\documentclass[aps,prl,twocolumn,sortedauthors,superscriptaddress,groupedaddress]{revtex4}
\usepackage{graphicx}
\usepackage{dcolumn}
\usepackage{bm}

\begin{document}

\preprint{hep-ex/xxxxx}
\title{Search for $\pi^0 \to \nu_{\mu}\bar\nu_{\mu}$ Decay in LSND}
\affiliation{University of California, Riverside, CA 92521}
\affiliation{University of California, San Diego, CA 92093}
\affiliation{University of California, Santa Barbara, CA 93106}
\affiliation{Embry Riddle Aeronautical University, Prescott, AZ 86301}
\affiliation{Los Alamos National Laboratory, Los Alamos, NM 87545}
\affiliation{Louisiana State University, Baton Rouge, LA 70803}
\affiliation{Southern University, Baton Rouge, LA 70813}
\affiliation{Temple University, Philadelphia, PA 19122}
\author{L.B. Auerbach}\affiliation{Temple University, Philadelphia, PA 19122}
\author{R.L. Burman}\affiliation{Los Alamos National Laboratory, Los Alamos, NM 87545}
\author{D.O. Caldwell}\affiliation{University of California, Santa Barbara, CA 93106}
\author{E.D. Church}\altaffiliation{Present address: Prediction Company LLC, Sante Fe, NM 87505}\affiliation{University of California, Riverside, CA 92521}
\author{A.K. Cochran}\affiliation{Southern University, Baton Rouge, LA 70813}
\author{J.B. Donahue}\affiliation{Los Alamos National Laboratory, Los Alamos, NM
87545}
\author{A.R. Fazely}\email{fazely@phys.subr.edu}\affiliation{Southern University, Baton Rouge, LA 70813}
\author{G.T. Garvey}\affiliation{Los Alamos National Laboratory, Los Alamos, NM
87545}
\author{R.M. Gunasingha}\affiliation{Southern University, Baton Rouge, LA 70813}
\author{R.L. Imlay}\affiliation{Louisiana State University, Baton Rouge, LA 70803}
\author{G. Kahrimanis}\affiliation{Southern University, Baton Rouge, LA 70813}
\author{W.C. Louis}\affiliation{Los Alamos National Laboratory, Los Alamos, NM
87545}
\author{R. Majkic}\affiliation{Temple University, Philadelphia, PA 19122}
\author{A. Malik}\affiliation{Louisiana State University, Baton Rouge, LA 70803}
\author{K.L. McIlhany}\affiliation{University of California, Riverside, CA 92521}
\author{W.J. Metcalf}\affiliation{Louisiana State University, Baton Rouge, LA 70803}
\author{G.B. Mills}\affiliation{Los Alamos National Laboratory, Los Alamos, NM
87545}
\author{D. Rupnik}\affiliation{Southern University, Baton Rouge, LA 70813}
\author{V.D. Sandberg}\affiliation{Los Alamos National Laboratory, Los Alamos, NM 87545}
\author{D. Smith}\affiliation{Embry Riddle Aeronautical University, Prescott, AZ 86301}
\author{R.F. Somodi}\affiliation{Southern University, Baton Rouge, LA 70813}
\author{I. Stancu}\altaffiliation{Present address: Department of Physics and Astronomy, University of Alabama, Tuscaloosa, AL 35487}
\author{W.D. Strossman}\affiliation{University of California, Riverside, CA 92521}
\author{M. Sung}\affiliation{Louisiana State University, Baton Rouge, LA 70803}
\author{R. Tayloe}\altaffiliation{Present address: Department of Physics, Indiana University, Bloomington, IN 47405}\affiliation{Los Alamos National Laboratory, Los Alamos, NM
87545}
\author{G.J. VanDalen}\altaffiliation{Present address: Embry Riddle Aeronautical University, Prescott, AZ 86301}\affiliation{University of California, Riverside, CA 92521}
\author{W. Vernon}\affiliation{University of California, San Diego, CA 92093}
\author{N. Wadia}\affiliation{Louisiana State University, Baton Rouge, LA 70803}
\author{D.H. White}\affiliation{Los Alamos National Laboratory, Los Alamos, NM
87545}
\author{S. Yellin}\affiliation{University of California, Santa Barbara, CA 93106}
\author{H. Yi}\affiliation{Southern University, Baton Rouge, LA 70813}
\noaffiliation
\collaboration{LSND Collaboration}
\noaffiliation
\date{\today}

\begin{abstract}
We observe a net beam-excess of
$8.7 \pm 6.3$ (stat) $\pm 2.4$ (syst) events, above 160 MeV,
resulting from the charged-current reaction of
$\nu_{\mu}$ and/or $\bar\nu_{\mu}$ on C and H in the LSND detector.
No beam related muon background is expected in this energy regime.
Within an analysis framework of $\pi^0 \to \nu_{\mu}\bar\nu_{\mu}$, we set a
direct upper limit for this branching ratio of
${\Gamma}({\pi}^0{\rightarrow}{\nu}_{\mu}{\bar{\nu}_{\mu}})/{\Gamma}({\pi}^0{\to}all)<1.6 \times10^{-6}$ at $90\%$ confidence level.
\end{abstract}


\maketitle

The observation of the decay $\pi^0\to \nu\bar\nu$ would imply new
interesting physics. The pion has zero spin and odd intrinsic parity
(i.e. $J^P=0^-$), and it is represented by a wave function which has the
space transformation properties under inversion and rotation of a
pseudoscalar.
Momentum and angular
momentum conservation require that the decay $\nu$ and $\bar\nu$ possess
the same helicity. This decay provides an ideal laboratory to search for
the pseudoscalar (P)
weak interaction, because only the P interaction allows massless neutrinos and 
antineutrinos with
the same helicity in the final state.
Furthermore, if the neutrino mass is not zero and the
$Z^0$ couples to the right-handed neutrino with standard weak-interaction
strength, the branching ratio (BR) $B(\pi^0\to\nu_\mu \bar\nu_\mu)$
has a maximum value of
$5.4\times10^{-14}$ at the $\nu_\mu$ mass upper limit of $m(\nu_{\mu}) =
0.19$ MeV/c$^2$ \cite{pdg}.
It is noteworthy that a BR of $\approx 10^{-14}$ for
$\pi^0\to \nu\nu\gamma$ within the Standard Model is allowed.
Therefore, an observed BR $B(\pi^0\to\nu_\mu \bar\nu_\mu) >>
5 \times 10^{-14}$ would imply physics beyond the Standard Model.

To date limits have been set on $\pi^0 \to \nu\bar\nu$ derived from pion
production in beam stops.
An experimental upper limit, $\Gamma(\pi^0\to\nu_{\mu}\bar{\nu_{\mu}})/
\Gamma(\pi^0\to all) \leq 3.1 \times 10^{-6}$  at $90\%$ confidence level (CL), was set by
Hoffman \cite{hoff} by using the data from several beam-dump
experiments. Similar limits were obtained by Dorenbosch et al. \cite{dorn}.
An inclusive search for $\pi^0 \to \nu\bar\nu$ using
$K^+ \to \pi^+\pi^0$ has set an upper limit of $8.3 \times 10^{-7}
(90\%$ CL) \cite{atiy}.
 
The Liquid Scintillator Neutrino Detector (LSND) experiment was performed
at the Los Alamos Neutron Scattering Center (LANSCE).
An 800-MeV beam of protons incident on a predominantly water-copper target
provided the source for the $\pi^0$. This energy is well below kaon
production threshold, and almost
all pions are produced from p-nucleus interactions.
The detector,
located 30 m from the beam stop, contained 167 tons of dilute liquid
scintillator which served as the active target. The liquid scintillator was
viewed by 1220, 8"-diameter Hamamatsu photomultiplier tubes (PMTs) mounted
inside
the tank.  An active shield with 292, 5"-diameter PMTs vetoed cosmic
rays \cite{napt}. The read-out and the data acquisition system are
described in more detail elsewhere \cite{ath97}.

\begin{figure}
\resizebox{9.0cm}{9.0cm}{\includegraphics[0,0][567,567]{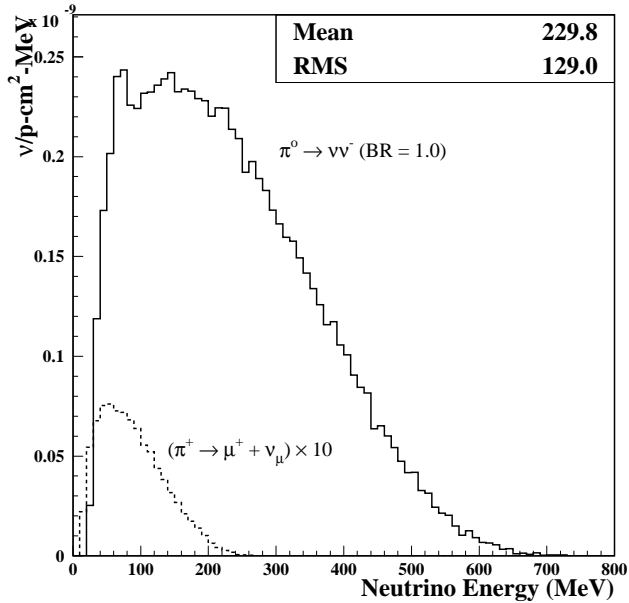}}
\caption{\label{neutrino}The neutrino energy spectrum resulting from the possible $\pi^0\to\nu_\mu \bar\nu_\mu$ decay at the beam stop and as seen at the 
center of the LSND detector. The dashed curve shows the energy spectrum of neutrinos from $\pi^+$ DIF.}
\end{figure}

The $\pi^0$ production rate was calculated using the LSND beam Monte Carlo (MC)
\cite{burm}. As the $\pi^0$ spectrum is expected to be very similar to that
of the $\pi^+$,
in these calculations the $\pi^+ \rightarrow e^+ \nu_e$ decay mode was used, where
a lifetime of $8.4 \times 10^{-16} s$ ($\pi^0$ lifetime) was assigned
to the $\pi^+$.
Assuming that all $\pi^0$'s decay to two neutrinos, the neutrino
flux is calculated to be $(6.5 \pm 1.6) \times 10^{14} \nu_{\mu}/cm^2$ at the 
center of the detector above the muon production threshold of 123 MeV for 
the entire running time of the LSND detector. The error estimate is based on 
the values quoted by Burman and Plischke \cite{burm1}. 
Fig. \ref{neutrino} shows the
neutrino spectrum resulting from the possible $\pi^0\to\nu_\mu \bar\nu_\mu$
decay at the center of the LSND detector.
Note that the peak between 70 and 80
MeV is due to multiple pion production \cite{burm1}. For comparison, Fig. 
\ref{neutrino}
also shows the $\nu_\mu$ spectrum from the usual
$\pi^{+} \to \mu^{+}\nu_{\mu}$ decay-in-flight (DIF) multiplied by 10.

The results presented here were obtained during the 1993-1998 LSND operation
periods. A total of 28,896 Coulombs of protons were incident on the beam stop.
The dominant interactions inside the LSND detector were
$\nu_{\mu} +^{12}C\to\mu^- +p +X$, $\bar\nu_{\mu} +^{12}C\to \mu^+ +n +X$ and
$\bar\nu_{\mu} +p\to\mu^+ +n$.
The flux-averaged cross sections for these reactions are estimated to be
$(3.0 \pm 0.6) {\times}10^{-38} cm^2$, $(0.60 \pm 0.12) {\times}10^{-38} cm^2$, and
$(0.20 \pm 0.01) {\times}10^{-38} cm^2$, respectively, from a Fermi-Gas model 
(FGM)
of Gaisser and O'Connell \cite{gais}. The estimated errors are from the 
uncertainties for the FGM calculations  
quoted by Vogel \cite{petr}. The ${\mu}$ in each of the
above reactions usually decays and produces a Michel electron.
The $\pi^0 \to \nu_{\mu}\bar\nu_{\mu}$ search
focused on identifying high-energy, muon-like beam excess
events in the energy range between 160 MeV and 600 MeV electron equivalent 
(MeVee).
The lower cut of 160 MeVee is obtained
because the end-point energy of muon neutrinos is about 280 MeV and the
muon production threshold is 123 MeV. This cut, therefore, insures that
beam-related background $\mu$ events are negligible \cite{auer}.
The upper energy cut was chosen because the end-point energy of the neutrinos
is about 730 MeV and the muon production
threshold is 123 MeV.

The majority of the triggers (99{\%}) were cosmic-ray
induced and were classified as
through-going muons,
electrons from stopping muon decays,
recoil protons from neutron collisions, and
$^{12}B_{g.s.}$ decay resulting from ${\mu}^-$ capture on $^{12}C$.
In order to minimize the background induced by cosmic rays,
a Michel electron was selected for the current event that satisfied
an energy cut
of $20<E<55$ MeV.
The reconstructed electron vertex was required to be inside a fiducial
volume 35 cm from
the face of the PMTs.
The particle identification parameter, ${\chi}_l^{'}$, from reference
\cite{agui} was used to select electrons and was required to satisfy
$-1.5<{\chi}_l{'}<0.5$, where the allowed range is chosen by
maximizing the
selection efficiency divided by the square root of the beam-off background.
A veto cut of less than 4 hits was applied to reduce cosmic rays.
Events with future activity within $9\,\mu\mbox{s}$ ($\approx 4 \mu $ lifetimes)
or with a bottom veto counter hit
were rejected in order to further eliminate cosmic-ray muon events.
Furthermore, no
veto hit was allowed within 30 ns of the trigger time. These cuts and their
associated efficiencies
for the selection of primary electrons are summarized in Table \ref{michel_eff}.
\begin{table}
\caption{Selection criteria and corresponding efficiency for the primary
Michel electron.}
\vspace{0.4 cm}
\begin{center}
\begin{tabular}{l l c} \hline\hline
Selection Criteria &         & Efficiency \\ \hline
Veto Live Time     &         & $0.76\,{\pm}\,0.02$\\ 
DAQ Live Time      &         & $0.96\,{\pm}\,0.02$ \\ \hline
                   & Analysis Efficiency   &   \\ \hline
Electron Energy    &$20\,<\,E\,<\,55\,\,MeV$      & $0.98\,{\pm}\,0.02$ \\ 
Fiducial Volume    & $D\,>\,35\,\,cm$         & $0.88\,{\pm}\,0.02$ \\ 
Particle ID        & $-1.5<\chi_{l}^{'}<0.5$ & $0.84\,{\pm}\,0.01$ \\ 
Shield and Crack Hits &$\,<\,4$             & $0.98\,{\pm}\,0.01$  \\ 
Future Time Gate   &${\Delta}t_{future}\,>\, 9\,\,\mu\mbox{s}$ & $0.99\,{\pm}\,0.01$\\ 
In-time Veto Gate  &${\Delta}t^{best}_{veto}\,>\,30\,\,ns$ & $0.97\,{\pm}\,0.01$\\\hline 
Total in-time Cuts &                    & $0.50\,{\pm}\,0.03$\\ \hline\hline
\end{tabular}
\end{center}
\label{michel_eff}
\end{table}
A second set of cuts was applied to isolate parent muons. These cuts are 
similar to those of reference [11]. Muons were
required to have a decay time of less than $9 \mu$s.
An energy
cut between 160 MeV and 600 MeV was chosen to eliminate possible muons from
beam-related $\pi^+$ DIF.
A spatial $\mu-e$ correlation distance cut of less than 100 cm was applied.
The muon vertex was required to be
within a fiducial volume of 35 cm from the face of the PMT.
A veto cut of less than 2 hits was applied to further reduce cosmic rays.
These cuts and their corresponding
efficiencies are
summarized in Table \ref{muon_eff}.

\begin{figure}
\resizebox{9.0cm}{9.0cm}{\includegraphics[0,0][567,567]{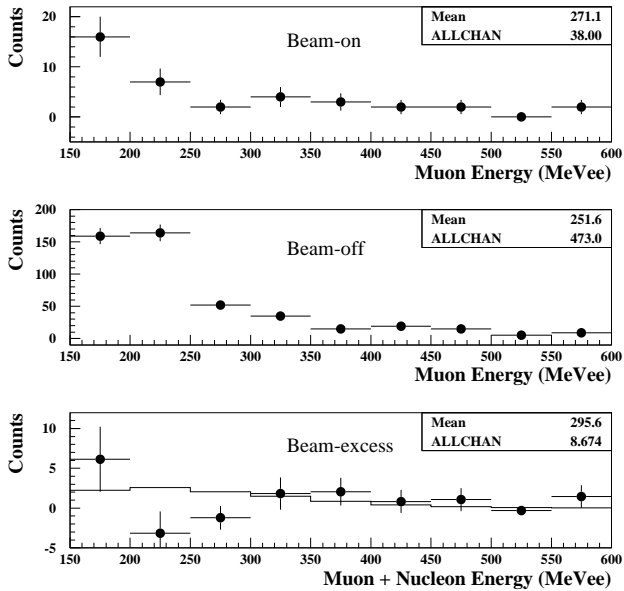}}
\caption{\label{excess_mu}The visible energy distribution for
beam-on, beam-off, and beam-excess events. The solid line represents
the MC calculation.}
\end{figure}

\begin{table}
\caption{Selection criteria and corresponding efficiency for the parent muon.}
\vspace{0.4 cm}
\begin{center}
\begin{tabular}{l l c} \hline\hline
Selection Criteria &         & Efficiency \\ \hline
Muon decay time    & $0.7<\tau_{\mu}<9\,\,{\mu}s$     & 0.74${\pm}$0.01 \\
Uncaptured $\mu^-$    & -     & 0.97${\pm}$0.01 \\ 
Past Energy  & $160\,\,MeV<E<600\,\,MeV$       & 0.64$\pm$0.01  \\ 
Spatial Correlation &$\,<\,100\,\,cm$        & 0.99${\pm}$0.01 \\
Fiducial Volume    & $\,>\,35\,\,cm$         & 0.75${\pm}$0.01 \\ 
Shield and Crack Hits &$\,<\,2$         &0.87${\pm}$0.01  \\ \hline
Total Past Cuts      &                  & 0.30${\pm}$0.01 \\ \hline\hline
\end{tabular}
\end{center}
\label{muon_eff}
\end{table}
The overall efficiency of these cuts
was $0.15 \pm 0.01$, which resulted in a total of 38 beam-on events and
473 beam-off events.
After applying the duty factor ratio of $0.062 \pm 0.005$,
the normalized number of beam-off events was $29.3 \pm 1.3 \pm 2.4$, resulting
in a beam-excess of $8.7 \pm 6.3 \pm 2.4$ events.
Fig. \ref{excess_mu} shows the visible energy distribution for
beam-on, beam-off, and beam-excess events. Also shown in the beam-excess 
figure
is the area normalized GEANT-3.21 MC calculation. The calculation uses the 
neutrino spectrum of Fig. 1 and generates muons and knock-out protons 
based on the Fermi-gas model of reference [9].      

\begin{figure}
\resizebox{9.0cm}{9.0cm}{\includegraphics[0,0][567,567]{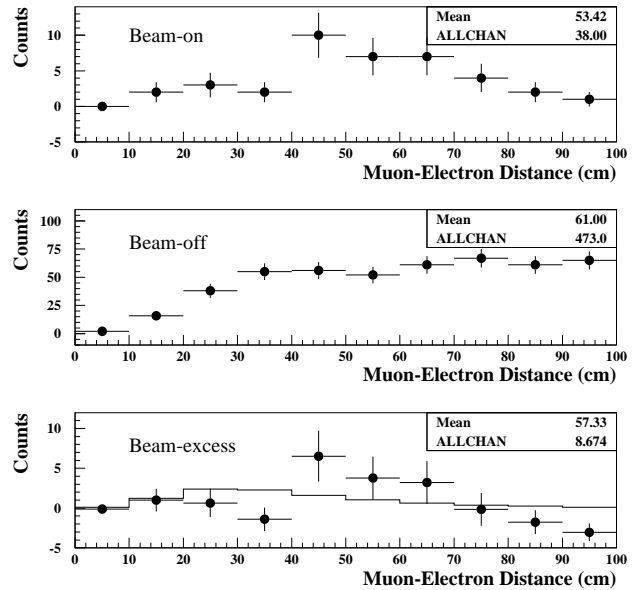}}
\caption{\label{excess_r}Distance between muon and electron for beam-on, beam-off and beam-excess events. The solid line represents the MC calculation.}
\end{figure}

\begin{figure}[h]
\resizebox{9.0cm}{9.0cm}{\includegraphics[0,0][567,567]{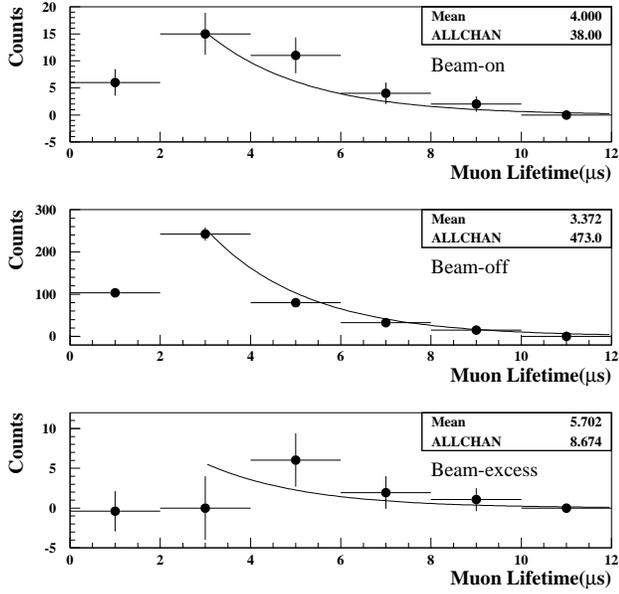}}
\caption{\label{excess_t}Muon lifetime for beam-on, beam-off and beam-excess 
events. The superimposed solid curves represent the expected 
$e^{-t/2.12\,\mu{s}}$ exponential decay of muons at rest.}
\end{figure}

Fig. \ref{excess_r} shows the spatial correlation between the muons and
the Michel electrons for the beam-off, beam-on, and beam-excess events.
The MC calculation
is shown as the solid line on the beam-excess histogram.
Fig. \ref{excess_t} shows
the time correlation between the muons and the Michel electrons for the
beam-on, the beam-off and the beam-excess events. The usual decay-at-rest muon 
lifetime curves, $e^{t/2.12\mu{s}}$ are superimposed for comparison. Note 
the choice of 2.12 $\mu\mbox{s}$ for muon lifetime is because the detected 
muons are a mixture of $\mu^-$ and $\mu^+$. 

The upper limit to the BR for the  $\pi^0 \to \nu_{\mu}\bar\nu_{\mu}$
decay is calculated using the values for
the neutrino reaction
cross section ($3.8 \pm 0.8) \times 10^{-38} cm^2$), the neutrino flux
($(6.5 \pm 1.6) \times 10^{14} \nu/cm^2$), the total number of target atoms
($(3.7 \pm 0.1) \times 10^{30}$),
and the overall efficiency ($0.15 \pm 0.01$).
If the small excess is due to
$\pi^0 \to \nu_{\mu}\bar\nu_{\mu}$ decay, then the corresponding BR is
${\Gamma}({\pi}^0{\to}{\nu}_{\mu}{\bar{\nu}_{\mu}})/{\Gamma}({\pi}^0{\to}all)=[6.4\pm 4.6 (stat)\pm 3.3 (syst)]\times10^{-7},$
which corresponds to an upper limit of
${\Gamma}({\pi}^0{\to}{\nu}_{\mu}{\bar{\nu}_{\mu}})/{\Gamma}({\pi}^0{\to}all)<1.6\times10^{-6}$ at $90\%$ CL. 

In summary, a beam-excess of $8.7 \pm 6.3$ (stat) $\pm 2.4$
(syst) events is observed above an
energy of 160 MeV, where no beam related background is expected.
Within the framework of
$\pi^0 \to \nu\bar\nu$, a direct upper limit for this BR of
$1.6 \times 10^{-6}$ at $90\%$ CL is obtained.

We are grateful to the administrative and technical
staff members of LANSCE for their support
during this experiment.
We also gratefully acknowledge Drs. Cyrus Hoffman, Peter Herczeg,
and Lon Chang Liu for enlightening discussions. This work was supported in 
part by the U.S. Department of Energy.


\end{document}